\pdfoutput=1
\documentclass[journal=cmatex,manuscript=article]{achemso}
\setkeys{acs}{articletitle = true}
\usepackage[usenames,dvipsnames]{xcolor}
\usepackage[bookmarks=false,colorlinks]{hyperref}
  \hypersetup{linkcolor=black,citecolor=black,filecolor=black,urlcolor=blue}
\usepackage[protrusion=true,expansion=true,final]{microtype}
\usepackage{multirow}
\usepackage{graphicx}
\usepackage{amssymb,amsmath,bm}
\usepackage{booktabs}

\title{Exploiting Colorimetry for Fidelity in Data Visualization}

\author{Michael J.\ Waters}
\affiliation{Department of Materials Science and Engineering, 
Northwestern University, Evanston, Illinois 60208, USA}

\author{Jessica M.\ Walker}
\affiliation{Department of Materials Science and Engineering, 
Northwestern University, Evanston, Illinois 60208, USA}

\author{Christopher T.\ Nelson}
\affiliation{Oak Ridge National Laboratory, Oak Ridge, Tennessee 37830, USA}

\author{Derk Joester}
\affiliation{Department of Materials Science and Engineering, 
Northwestern University, Evanston, Illinois 60208, USA}

\author{James M.\ Rondinelli}\email{jrondinelli@northwestern.edu}
\affiliation{Department of Materials Science and Engineering, 
Northwestern University, Evanston, Illinois 60208, USA}

\date{\today}
\begin{document}

\begin{abstract}
Advances in multimodal characterization methods fuel a generation of increasing immense hyper-dimensional datasets. 
Color mapping is employed for conveying higher dimensional data in 
two-dimensional (2D) representations for human consumption without relying on multiple projections. 
How one constructs these color maps, however, critically affects how accurately one perceives data.
For simple scalar fields, perceptually uniform color maps and color selection have been shown to improve data readability and interpretation across research fields. 
Here we review core concepts underlying the design of perceptually uniform color map 
and extend the concepts from scalar fields to two-dimensional vector fields and three-component composition fields frequently found in materials-chemistry research to 
enable high-fidelity visualization. 
We develop the software tools PAPUC and CMPUC to enable researchers to utilize these colorimetry principles and 
employ perceptually uniform color spaces for rigorously meaningful color mapping of higher dimensional data representations.
Last, we demonstrate  how these approaches 
deliver immediate improvements in data readability and interpretation in microscopies and spectroscopies routinely used in discerning  materials structure, chemistry, and properties.  
\end{abstract}

\maketitle

\section{Introduction}
Scientific, engineering, and medical research are expensive endeavors and 
include studies ranging from fundamental materials chemistry to materials deployment.
This broad community of materials chemists and its stakeholders seek to ensure that 
complex data sets generated in this enterprise are accurately perceived; 
however, researchers have become increasingly aware of how 2D representation, in the form of color maps or heat maps, which define the conversion of raw data to colors on a screen or in print, can be poorly constructed to obscure or over-represent data \cite{ArteryMap, wong2010, wong2011-2, gehlenborg2012-1, gehlenborg2012-2}.
Furthermore, 2-3\% of the human population suffers from deuteranomaly---the most 
common form of red-green blindness. \cite{ColorblindnessWorldwide, wong2011-1}
Which limits their ability to differentiate hues in the red-orange-yellow-green palette. 
Color maps, therefore, should be inclusive for broad accessibility while providing a faithful representation.
The current state of scientific data visualization uses Red-Green-Blue (RGB) values directly 
when creating visual representations of experimental or computationally acquired data. 
RGB color spaces, are conceptually easy to comprehend as 3D coordinate systems that
approximate the tristimulus values, \textit{i.e.} the stimulation of the red, green, and blue receptors in the human eye.
Although there are multiple RGB standards and high-dynamic range (HDR) standards will improve their gamut, the fallback choice remains the sRGB  standard.\cite{sRGB_standard} 
For data visualization, there are many issues with using the sRGB color space---some disadvantages are described below. 
The main limitation being that it is not a perceptually uniform color space; 
if one specifies uniformly spaced hue values, the corresponding colors will not 
disperse linearly to the human eye.
Thus, the aspirational goal of optimal color usage for achieving fidelity in 
data visualization requires moving away from sRGB to the more appropriate perceptually uniform colors spaces (UCSs).

\begin{figure}[t]
\centering
\includegraphics[width=0.30\textwidth,clip]{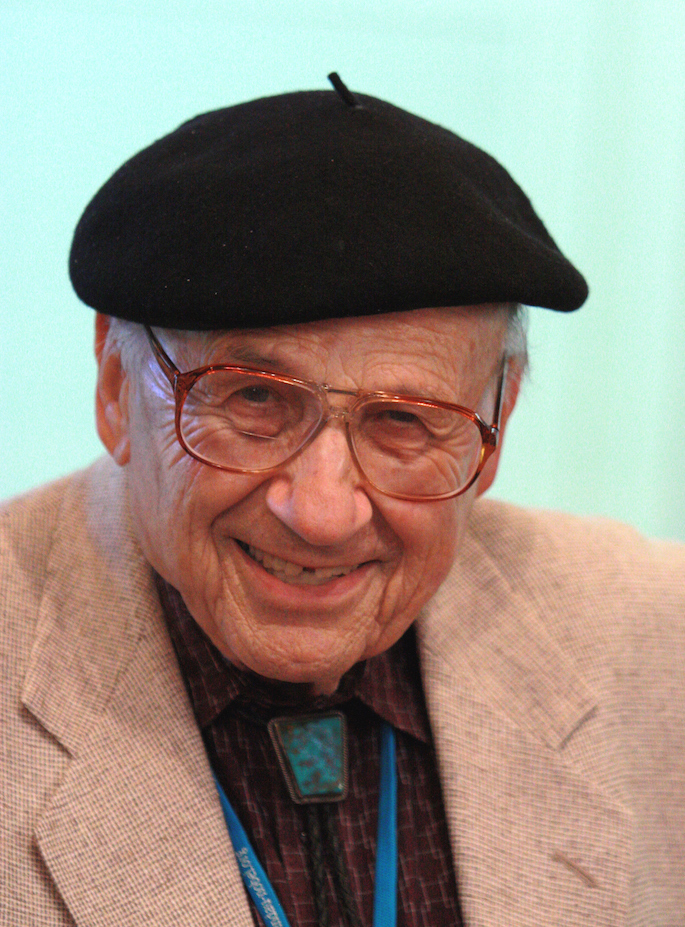}
\includegraphics[width=0.30\textwidth,clip]{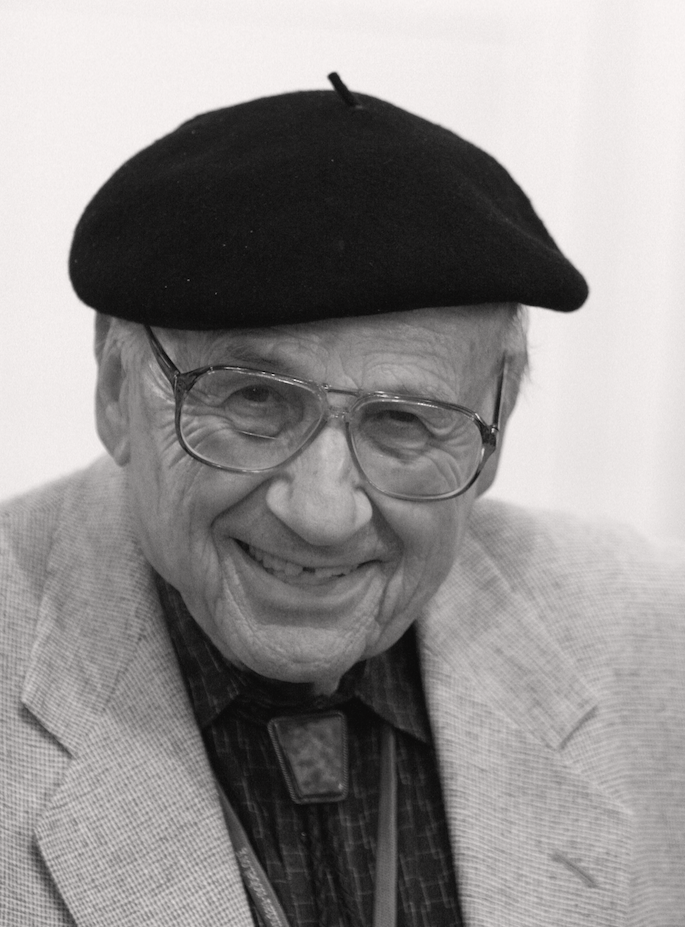}
\includegraphics[width=0.30\textwidth,clip]{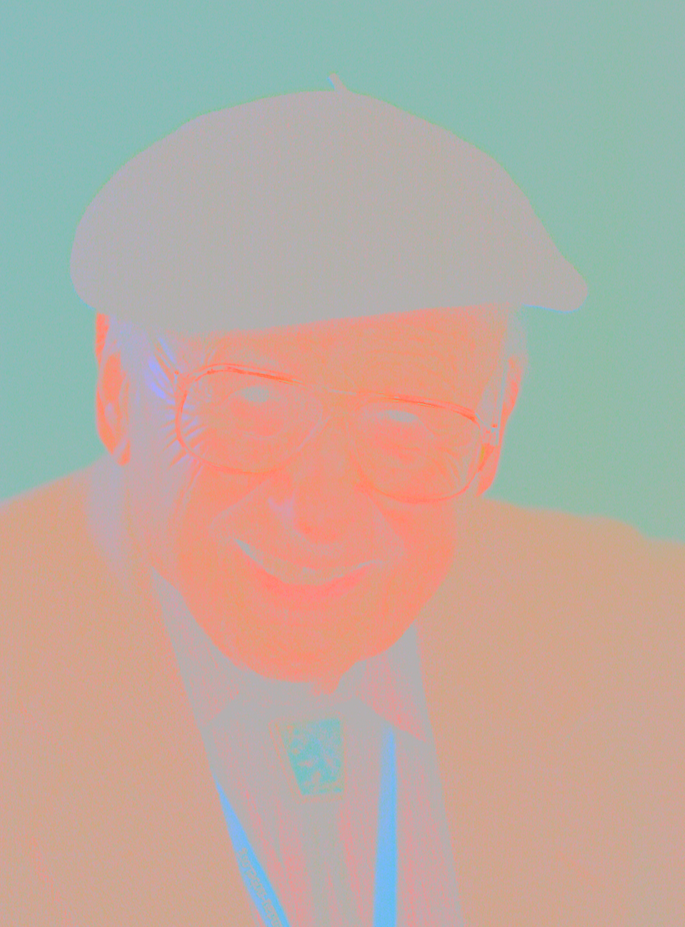}
\caption{\label{fig:WalterKohn} (Left) A photograph of Nobel Laureate Walter Kohn and two modifications of the image where (center) $a^{*},  b^{*} \rightarrow 0$ yielding gray scale and (right) $L^{*} \rightarrow 75$ removing lightness information. Lightness information is generally much easier to process. Images adapted from Ref.\ \cite{Walter_Kohn}.
}
\end{figure}

Today, large selections of off-the-shelf, aesthetically pleasing, perceptually uniform color maps for scalar fields are both available and can be readily designed with available tools. 
They are designed such that the 
Euclidian distance between any two color coordinates are a perceptual distance. 
They are, in essence, models of how humans process color information. 
\cite{ColorAppearanceInCortex}
As a demonstration, we provide an original (\autoref{fig:WalterKohn}a) and variations (\autoref{fig:WalterKohn}b and c) of a photograph of Nobel Laureate Walter Kohn (\autoref{fig:WalterKohn}), owing to the immense number of materials-chemistry publications that include electronic structure calculations (and derived properties) from density-functional theory.
The modified photographs illustrate how colorimetry choices affect data (image) perception.
In \autoref{fig:WalterKohn}b, we remove the hue and saturation data whereas we 
remove the lightness data in \autoref{fig:WalterKohn}c. 
These transformations are trivial in UCS.
\autoref{fig:WalterKohn} also demonstrates how lightness information alone is easy to comprehend, while hue and saturation data alone are more difficult.
The importance for data representation lies in that perceived changes between colors can be mapped intuitively in a linear fashion to the changes in data they represent. 
\begin{figure}[t]
\centering
\includegraphics[width=0.45\textwidth,clip]{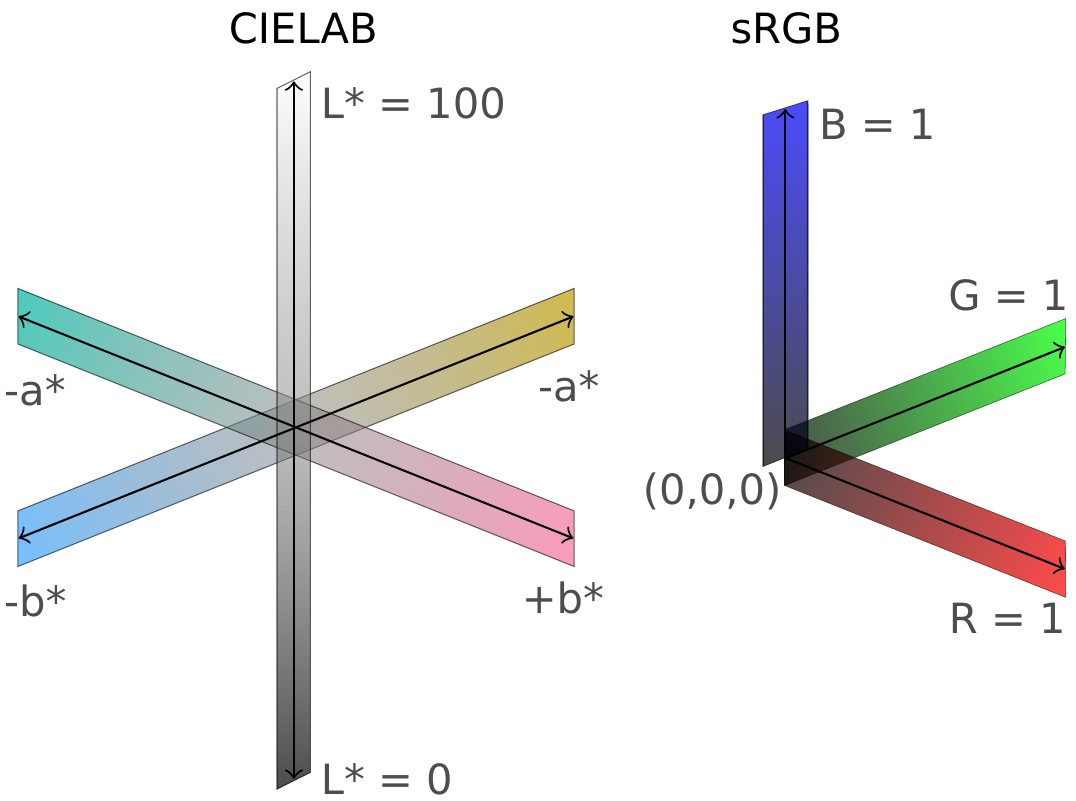}
\caption{\label{fig:cielab} The (left) CIELAB and (right) sRGB color space axes and their hues. The sRGB color space forms a unit cube where the origin is black, the cube vertices on the axes are red, green, and blue primaries, and white is on furthest vertex from origin. 
}
\end{figure}
The CAM02-UCS and CAM16-UCS colorspaces\cite{CIECAM,Li2017} represent the current state-of-the-art 
and should be utilized for 
diverse scientific visualizations. 
However the simpler and more widely used CIELAB is sufficient for demonstration purposes while being easily adoptable. 
In CIELAB, $L^{*}$ is lightness and $a^{*}$ and $b^{*}$ are color channels, where lightness ranges from 0 to 100 (unitless) and the black-gray-white trajectory 
lies along $a ^{*} , b ^{*} = 0$ (\autoref{fig:cielab}).
The perceptual distance 
$\Delta E = 
[\left ( \Delta L ^{*} \right )^2 + \left ( \Delta a ^{*} \right )^2 + \left ( \Delta b ^{*} \right )^2 ]^{1/2}$
provides a measure of how far apart, or rather, how much contrast 
exists between two colors. 
Thus, one can infer the utility of using UCSs 
for data representation so as not to obfuscate variations in color contrast with 
variations in data (\textit{e.g.}, scalar magnitudes). 
For color mapping data, we further link the color contrast to the 
perceivable data contrast using the derivative of the perceptual distance 
with respect to the data, $X$, as $\partial(\Delta E) / \partial X$.
Having a constant perceptual derivative is what defines a perceptually uniform color map.  
Since most people only recognize a $\Delta E=2.3$,\cite{DigitalColorHandbook} the 
perceptual derivative allows us to compute the minimum perceivable change in the data 
with a certain user- or instrument-specified color mapping.
If a non-perceptually uniform mapping is made, then the value of the derivative will 
vary with the data values. . 
For simple scalar fields, perceptually uniform color mapping can be as simple as 
linear interpolation between two points in a UCS.   
However, to increase the perceptual derivative and thus the data contrast without 
sacrificing perceptually uniformity, a helical path along the lightness axis is often used. 
Another benefit of this path is that it is linear luminant---
the lightness derivative is constant. 
This feature ensures good readability for those with color-blindness or 
in gray scale, since hue differentiation is not required. 
An alternative simple way to address the deuteranomaly is to ensure that 
red and green do not appear next to each other in the color space.

\begin{figure}[t]
\centering
\includegraphics[width=0.45\textwidth,clip]{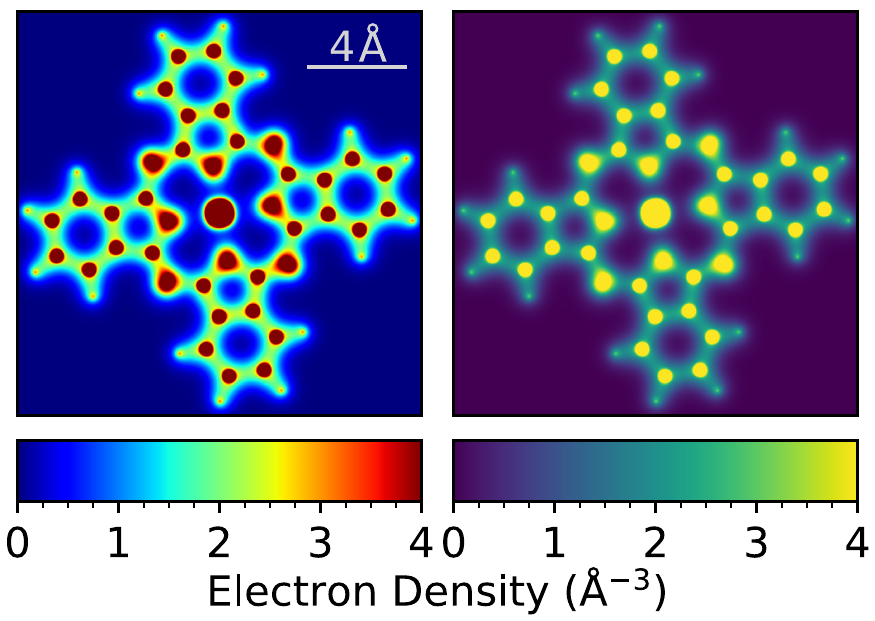}
\caption{\label{fig:edensity} Electron density slice through a copper phthalocyanine molecule, computed with \texttt{GPAW}\cite{gpaw}, is colormapped with the ubiquitous \emph{jet} colormap (left) and the perceptually uniform and linear luminant \emph{viridis} colormap of  \texttt{Matplotlib} (right).
}
\end{figure}

Recently,  Matplotlib, MATLAB\textsuperscript{\textregistered}, and other scientific and engineering 
software changed their default color maps to be perceptually uniform, linear luminant, and 
compatible with red-green color blindness. 
Nonetheless, \textit{ad hoc} and/or purely aesthetically designed color maps
remain pervasive in figures rendered for both 
electronic and print publication.
In contrast, well-designed, perceptually uniform color maps exhibit uniform contrast both in lightness and hue across the whole color map 
while avoiding the red-green section of the spectrum. 
These maps are excellent for simple scalar fields such as height or density data (\autoref{fig:edensity}). 
For amplitude data, which has both positive and negative values, there are perceptually uniform 
diverging color maps \cite{CET}. 
Likewise, for cyclic fields, such as phase angle, there are perceptually uniform cyclic colormaps \cite{CET}. 

Here we describe two tools to facilitate the use of UCSs and demonstrate their use with real 
materials data. 
First, we demonstrate the Phase-Amplitude Perceptually Uniform Colormap Designer (PAPUC) tool, which is implemented as a small library for creating perceptually uniform color mappings of vector fields where vector orientations/phase angles are mapped onto hue and vector magnitudes/amplitudes are mapped onto lightness. 
Next, we demonstrate the Composition Mapping within Perceptually Uniform Colorspaces (CMPUC) tool for performing perceptually uniform color mappings of three component composition fields. 
A common use-case for this is energy-dispersive x-ray spectroscopy (EDS) mapping of elements in multicomponent materials where traditionally red, green, and blue are used to represent different comprising elements.
We then showcase the use of these tools, both written in Python and are integrated into the \texttt{SciPy}/\texttt{NumPy}/\texttt{Matplotlib} ecosystem,\cite{SciPy-NMeth,numpy,MPL}
for a 2D vector field in a complex oxide heterostructure and a multicomponent composition mapping in the common hydroxyapatite biomineral. 
The tools are  freely available at 
Refs.\ \cite{zenodo.3688502,zenodo.3688508} 
and use \texttt{colorspacious} \cite{colorspacious} for back-end color space conversions. 
The \texttt{Colour-Science}\cite{colour} package will likely be added as another back-end option in the future. 
The concepts we demonstrate are useful to data acquired
in various scientific and technological fields; both of the aforementioned examples use data collected with transmission electron microscopy (TEM), which is a broadly
utilized multimodal characterization tool in materials chemistry.

\section{Methods}

\subsection{Vector Fields}

\begin{figure}
\centering
\includegraphics[width=0.45\textwidth,clip]{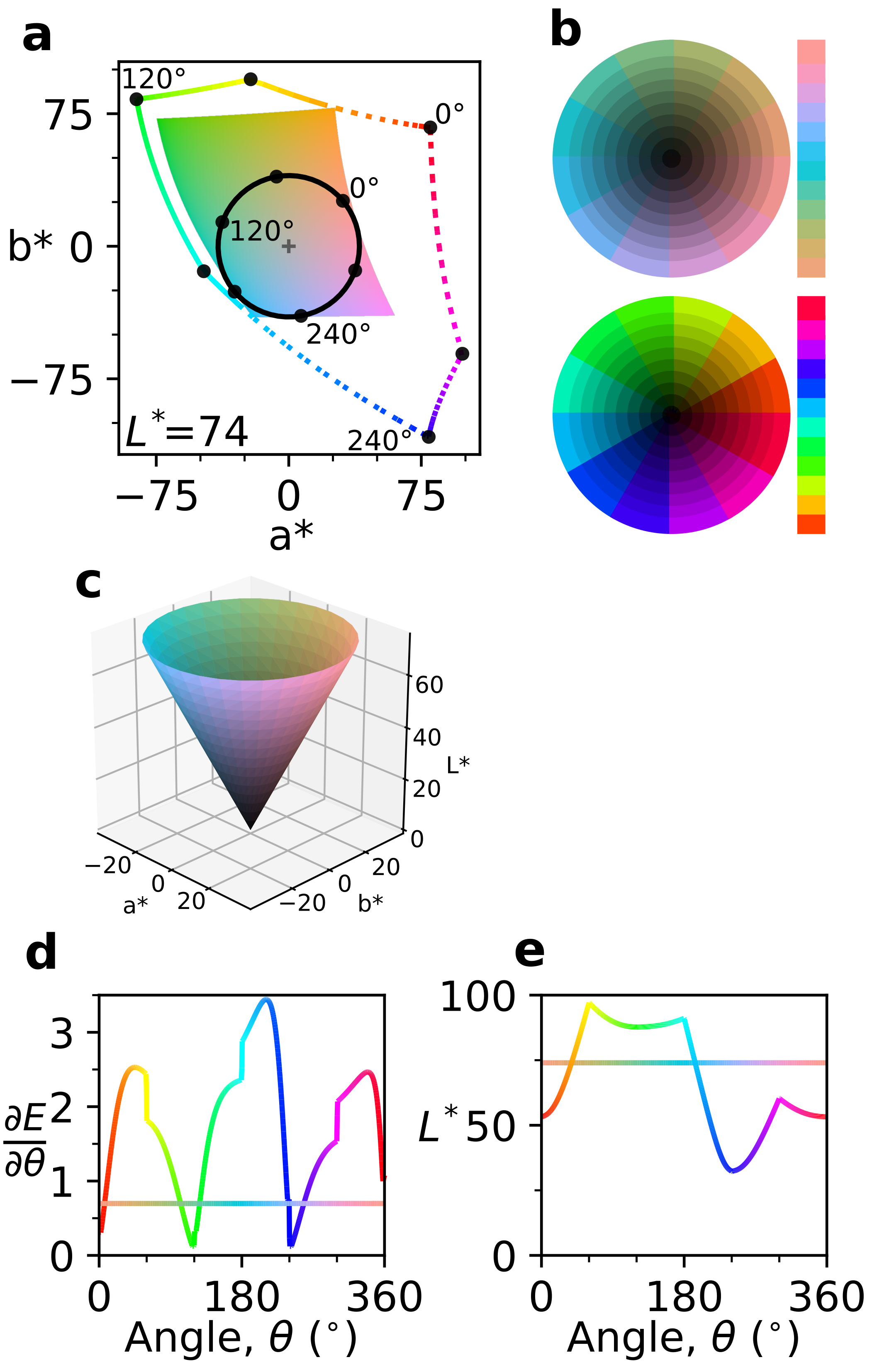}
\caption{\label{fig:polar1} (a) A slice of the CIELAB color space at $L ^{*}=74$. The black circle of radius 40 forms the base of the cone used in our perceptually uniform color mapping. The region that can be represented in the sRGB gamut is colored to match. The outermost line is the path that the outer ring of HSV takes through CIELAB space and is shown with is solid line where $L ^{*} \geq 74$ and broken elsewhere. HSV passes through the sRGB primaries at $H=0^{\circ}, 120^{\circ}, 240^{\circ}$, and secondaries at $H=60^{\circ}, 180^{\circ}, 300^{\circ}$ if $V=1$. (b) The perceptually uniform color wheel is shown with the  origin hue shifted to match the HSV color wheel. Vector magnitude and angle are shown in 10\% and $30^{\circ}$ increments, respectively, (c) The cone in the uniform color space upon which vector angle and magnitude are mapped. (d) The perceptual derivatives of the HSV and perceptually uniform color wheels with respect to angle. (e) Lightness as a function of vector angle for both color wheels.}
\end{figure}

In representing 2D vector fields, it is common to use the HSV (hue, saturation, value) 
color space, which is a mathematically simple transformation of sRGB. 
The value of each pixel, \emph{i.e.}, brightness and hue, are determined by magnitude and angle, respectively.
Here we use the CIELAB color space to map the direction and magnitude of the vectors (\autoref{fig:polar1}).
Using the perceptually uniform representation, as implemented in PAPUC, the vector angle $\theta$, is mapped to the hue angle, $h^{\circ}$, in the $a ^{*}$-$b ^{*}$ plane (\autoref{fig:polar1}a) and lightness is scaled to vector magnitude so that the perceptually uniform color wheel has a conceptually familiar layout as the HSV color wheel (\autoref{fig:polar1}b). When visualized in CIELAB colorspace, input vector data are mapped on to
the surface of an inverted cone with an axis along $L^{*}$, cone tip at $L^{*}=0$, 
cone base at $L^{*}=74$, and radius of 40 which can be seen in \autoref{fig:polar1}c.  (More details may be found in Ref.\  \cite{zenodo.3688502}.). 

The angle contrast is quantified by the perceptual derivative with respect to angle, 
$\partial \Delta E / \partial \theta$, at the maximum magnitude the HSV perceptual derivative varies by an order of magnitude and is discontinuous (\autoref{fig:polar1}d). 
Further complicating the HSV representation, lightness varies wildly with respect to the angle 
$\theta$, which leads some angles to have inherently higher perceived magnitude (\autoref{fig:polar1}e). 
This is a result of HSV passing through the sRGB primaries, where red and green are 1.65X and 2.72X 
lighter than blue. 
The perceptual derivative discontinuities occur at the sRGB primaries and their binary 
combinations of cyan, magenta, and yellow. 
Assuming a just noticeable difference of $\Delta E=2.3$ \cite{DigitalColorHandbook}, 
our proposed mapping permits a minimum discernible angle of 3.3$^{\circ}$. 

\begin{figure}[t]
\centering
\includegraphics[width=0.45\textwidth,clip]{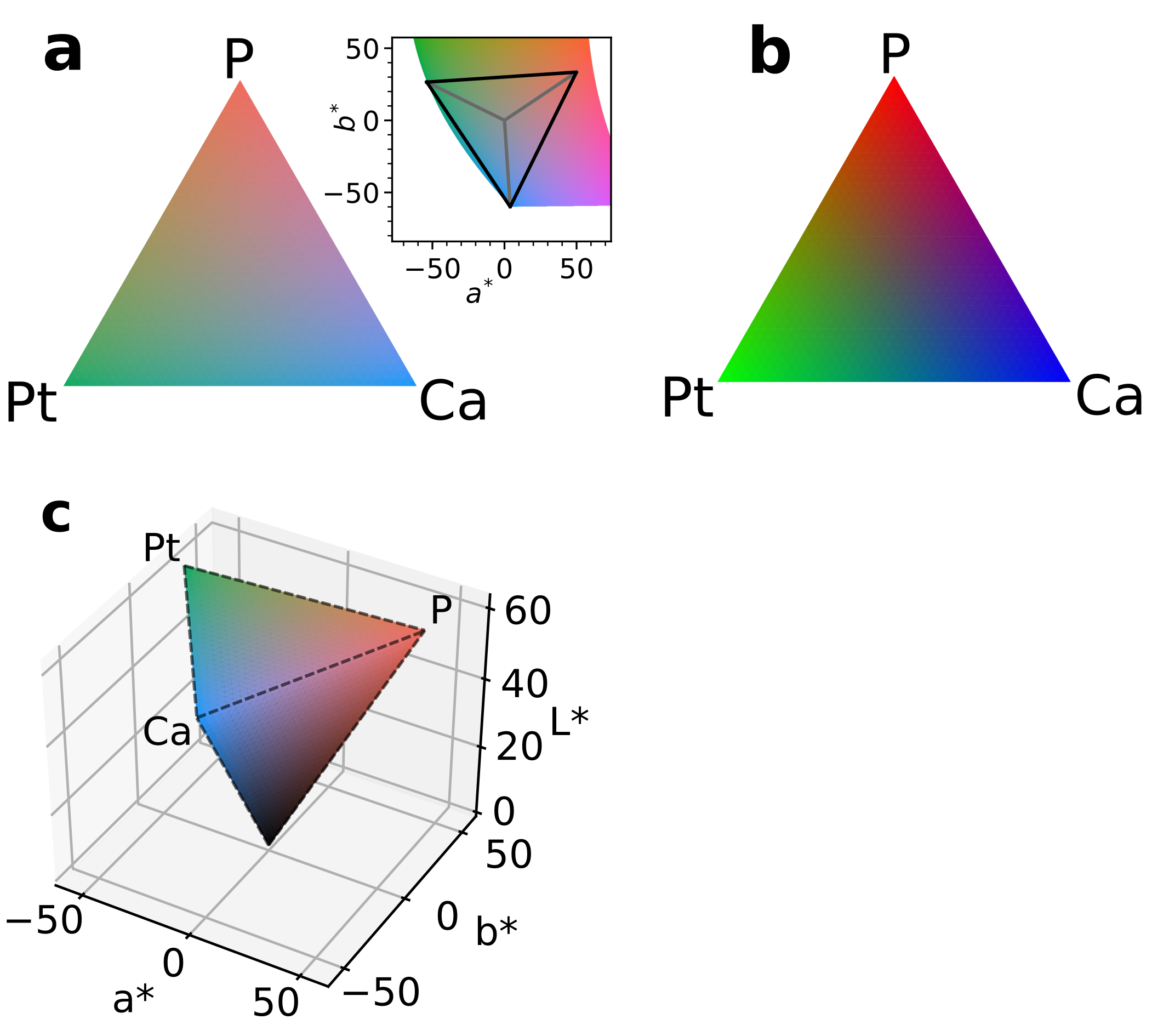}
\caption{A chemical color mapping where red, green, and blue represent three different elements (P, Pt, and Ca, respectively) where (a) is perceptually uniform and the three color points form an equilateral mixing triangle in the $a ^{*}$-$b ^{*}$ plane while (b) depicts a linear mixing triangle of the sRGB primaries, \textit{i.e.}, permutations of (1,0,0). (c) The inverted trigonal pyramid formed by perceptually uniform chemical mapping in the uniform color space.}
\label{fig:composition1}
\end{figure}

\subsection{Compositional Mapping}
In mapping elemental composition  in materials,  typically the three sRGB primaries or their binary pairs,  
cyan, magenta, and yellow, are used to represent the measured concentrations of three components.
In our approach implemented in CMPUC, we  scale lightness, $L^*$, to the total intensity (concentration).
For the relative composition, three color points are chosen such that they are perceptually equidistant and 
centered at white ($a ^{*} , b ^{*} = 0$) forming an equilateral triangle (\autoref{fig:composition1}a).
Ideally, the equilateral triangle in the $a^*$-$b^*$ plane should be as large as 
possible so that the smallest changes in composition are perceivable. 
Thus, we use the largest white centered triangle that fits into the sRGB color gamut; 
which occurs at $L^{*}=61.50$, circumradius of 60.14, and hue angle of 33.73$^{\circ}$. 
The circumradius of the triangle is scaled to the total intensity since the sRGB gamut 
converges to ($a ^{*} , b ^{*} = 0$) as $L^{*}$ approaches zero. 
The overall mapping in our CMPUC code is essentially an inverted trigonal pyramid which has a triangular base where each base vertex corresponds a maximum signal that is purely of one component (\autoref{fig:composition1}c). 

In traditional RGB mapping, the foremost issue with the approach is that the primaries exhibit different lightness values. 
The lightness variation is so large that the component represented by green will be perceived as 
nearly 3X more concentrated than the component represented by blue. 
Although the sRGB primaries have maximal perceptual distance and thus, compositional distinguishability, 
they are not perceptually equidistant, such that there are more perceivable differences between blue and green 
than either of the other pairs. Additionally, linear mixing between them yields low contrast near pure compositions.

\section{Use-cases}

\subsection{Application to vector-field visualizations}
First, we show how perceptually uniform color mappings
give immediate improvements in data readability and interpretation using our software tool PAPUC
for a vector field consisting of nanoscale vortex patterns of electrical polarization in a PbTiO$_{3}$/SrTiO$_{3}$ superlattice measured from atomic-scale Scanning Transmission Electron Microscopy (STEM).\cite{PolarVortex}  
The rotational structure and magnitude differences of this vector field provide an ideal test for the interpretability of color mapping methods as they rotationally and radially traverse the entire color space.

\begin{figure}[t]
\centering
\includegraphics[width=0.45\textwidth,clip]{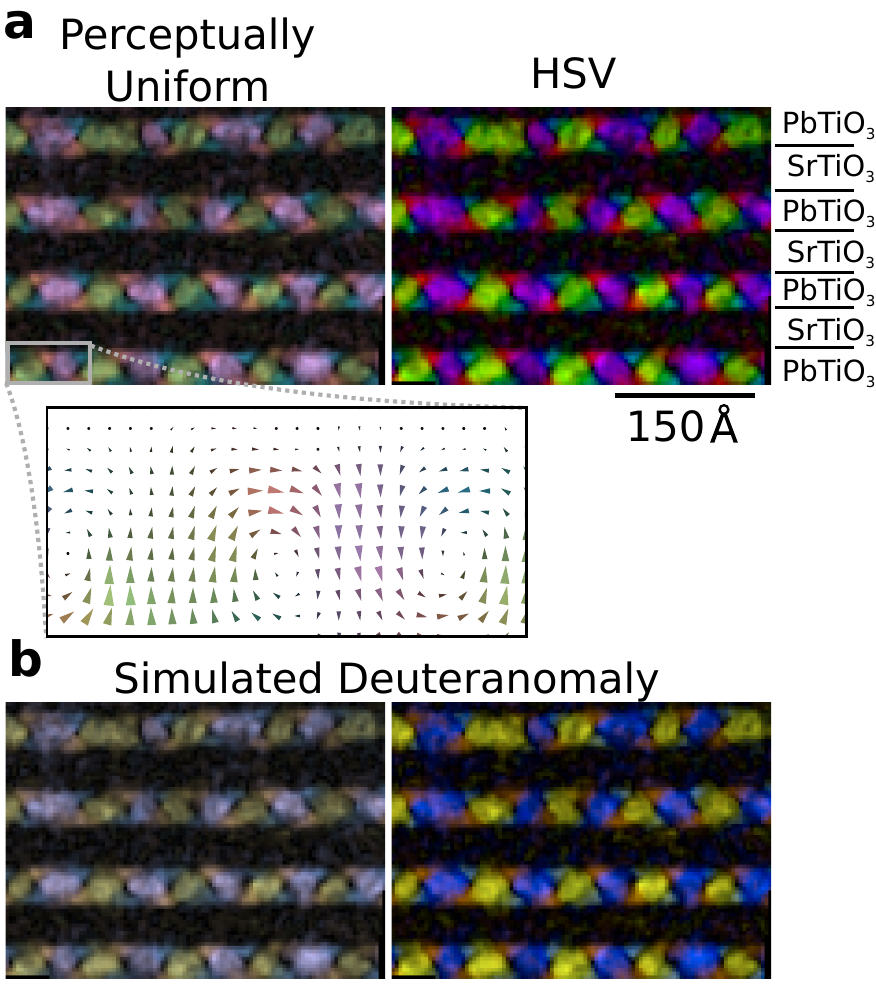}
\caption{\label{fig:polar2} (a) Ferroelectric polarization domains in a super lattice of PbTiO$_{3}$/SrTiO$_{3}$ as visualized by angle as hue and intensity as lightness with our perceptually uniform mapping and with HSV. (b) The same two images simulated with the deuteranomaly. Pixel spacing of 3.94 \AA/pixel.}
\end{figure}

\autoref{fig:polar2} presents a dataset containing ferroelectric PbTiO$_{3}$ layers with a spontaneous polarization that form a series of vortices when alternated with non-polar SrTiO$_{3}$ layers (see inset). The electrical polarization vectors are mapped with our perceptually uniform representation (left) and the common HSV (right) representation.
Most notably, the apparent magnitude of the vector is well preserved in the perceptually mapping but not in the HSV mapping. 
The vectors principally along the Y axis render in opposing extremes of lightness, either low lightness blue (-Y) or a high lightness yellow (+Y) for a given magnitude, making accurate visual inspection challenging. For the same reason, the rotation of the vector field is also better preserved, such as the low magnitude “hole” at the vortex centers and a higher rotation gradient along preferential planes creating a slight “X” pattern. Finally, there is less hue distortion when using the perceptually uniform mapping under simulated deuteranomaly.

\begin{figure}[t]
\centering
\includegraphics[width=0.45\textwidth,clip]{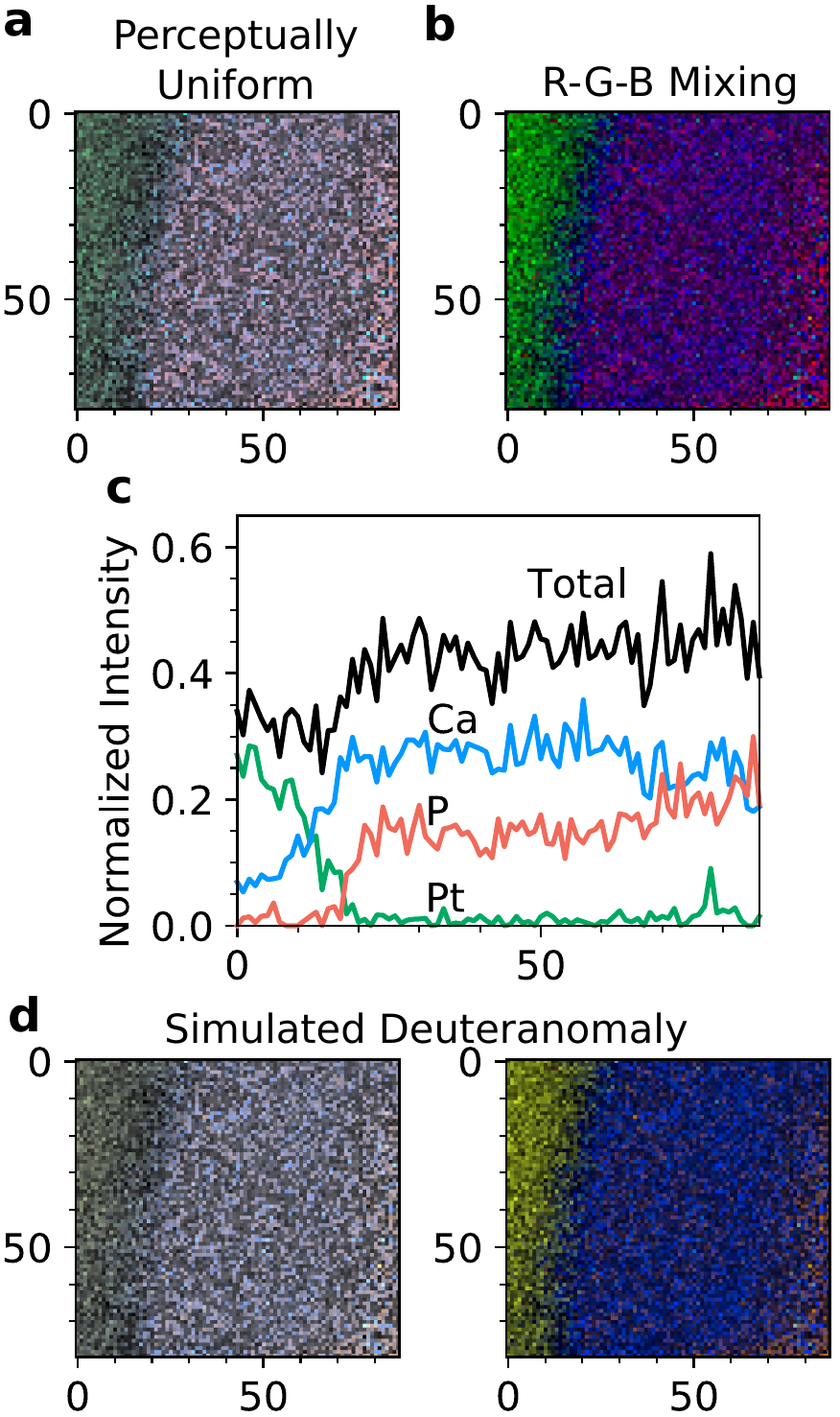}
\caption{Energy-dispersive X-ray spectroscopic chemical mapping of platinum deposited on hydroxyapatite. Red, green, and blue represent P, Pt, and Ca signals, respectively where (a) is perceptually uniform and (b) uses linear mixing of the sRGB primaries, \textit{i.e.} permutations of (1,0,0). Lengths are given in pixels. (c) The average signal for the bottom 20 horizontal line scans. (d) The chemically mapped images with simulated deuteranomaly.}
\label{fig:composition2}
\end{figure}

\subsection{Application to compositional maps}

Next, we examine the application of our perceptually uniform composition mapping using CMPUC on TEM EDS composition data collected from a cross section sample of  
hydroxyapatite deposited with platinum  (\autoref{fig:composition2}). 
As before,  we compare the perceptually uniform visualization (\autoref{fig:composition2}a) to the traditional mapping method, which in this case is R-G-B mixing (\autoref{fig:composition2}b). When pairing colors to chemical elements in either mapping (\autoref{fig:composition1}), we advocate for match-ups that minimizes spatial overlap between red and green, to aid those with red-green color blindness. 

Using the R-G-B mixing color map, the platinum-rich region on the left side of \autoref{fig:composition2}b appears to have higher signal; however, the total signal is lowest in the platinum-rich region as shown in \autoref{fig:composition2}e.
This inconsistency in the R-G-B mixing color map shows how our perceptually uniform color map yields a more accurate  
representation of the total signal while maintaining good compositional distinguishability as seen in \autoref{fig:composition2}c. 
One might argue that the small amount of platinum signal visible on the right side of 
\autoref{fig:composition2}b,c is not noticeable within our perceptually uniform color mappping; however, if the elemental 
color pairings were permuted and Pt was represented as blue, the small platinum signal would not be visible in R-G-B mixing color mapping.
To that end, if emphasizing small signals in a quantitative manner is important, 
it is best to separate elemental signals into separate figure panels and use a scalar colormap such as \emph{viridis} as it provides $\approx$30\% more contrast than using a black-green colormap.

\section{Conclusions}
Using colorimetry to design better---in a scientific fidelity rather than aesthetic sense---data representations for authentic perception is gaining substantial traction. 
Yet many of the current off-the-shelf tools, may not cover the more complex data representations with perceptually uniform use of color. 
The two examples and their implementations in PAPUC and CMPUC we highlighted are only marginally more complex than normal data post-processing as the software implementation of $L^*\,a^*\,b^*$-sRGB color conversion is trivial.
Yet for minimal extra effort, we demonstrate significant improvements in interpretability over the common HSV color mapping of 2D vector fields. Most importantly, the fixed lightness-magnitude mapping no longer hides vectors mapped to blue and we can ensure uniform angle contrast.
Last, within three-component chemical mapping, we demonstrate a fair lightness-concentration representation for all three components.   
We hope, above all else, that our work encourages others to use perceptually uniform color mappings since the community of scientific researchers all strive for the accurate conveyance of data.

\acknowledgement
M.J.W.\ was supported by the Office of Naval Research (ONR) (Contract No.\ N00014-16-1-2280).
J.M.R.\ acknowledges support from an Alfred P.\ Sloan Foundation fellowship (Grant No.\ FG-2016-6469).
C.T.N. was supported by the U.S. Department of Energy (DOE), Office of Science, Basic Energy Sciences, Materials Sciences and Engineering Division.
We thank Prof.\ R.\ Ramesh for his feedback and financial support of C.T.N.\ during initial data collection.
Human enamel sample data presented in \autoref{fig:composition2} were provided by the Procter \& Gamble Company.

\bibliography{color_map_references}

\newpage

\centering Table of Contents Graphic
\begin{figure}[t]
\centering
\includegraphics[width=6.9cm]{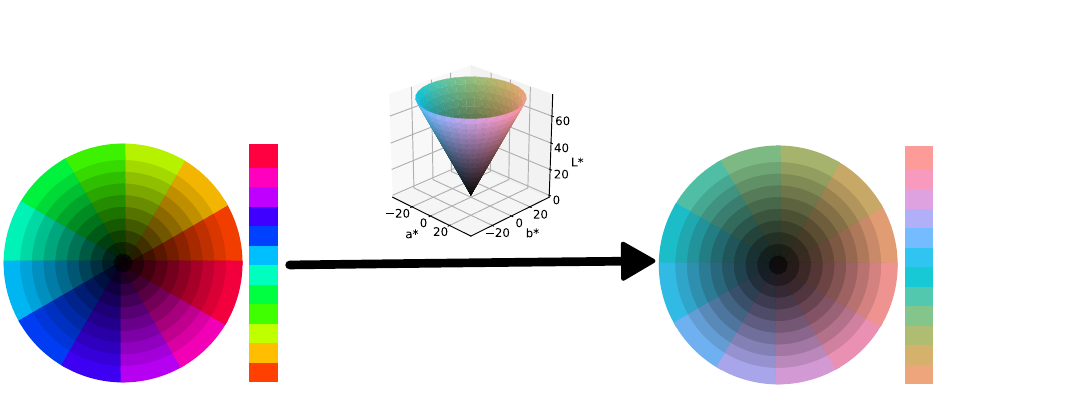} 
\label{fig:TOC}
\end{figure}

\end{document}